# Spinodal Surface Fluctuations on Polymer Films

Y. J. Wang and Ophelia K. C. Tsui*

*Department of Physics and Institute of Nano Science and Technology, Hong Kong University of Science and Technology, Clear Water Bay, Hong Kong, China.*

We study the temporal growth pattern of surface fluctuations on a series of spinodally unstable polymer films where the degree of instability is controlled by the film thickness. For films in the deep spinodal region, the growth rate function of the surface modes as a function of the wavevector $q$, $\Gamma(q)$, fits well to the mean-field theory. As the film thickness is increased and the film instability decreases, the mean-field theory demonstrates marked disagreement with experiment, notwithstanding provision of the known corrections for random thermal noise. We show that the deviations arise from large-amplitude fluctuations induced by homogeneous nucleation, which has not been considered in conventional treatment of thermal noise.

According to phase transition theories, off-critical systems with dimensionalities below the upper critical value of 6 would have the mean-field spinodal smeared and the mean-field predicted phase properties invalidated by thermal fluctuations near the spinodal [1-3]. While non-mean-field behaviors are expected on both sides of the spinodal for these cases, the majority of studies have focused on the effect of thermal fluctuations on the limit of metastability. No systematic experiment dedicated to investigate how the phase behavior may be altered on the unstable side has been done. Here, we examine the effect of thermal fluctuations on the spinodal growth of the order parameter in a 2-dimensional (2D, well below the upper critical dimension) unstable system. The system studied is polystyrene (PS) deposited on Si covered with a 106 nm thick $SiO_x$. In this paper, results of PS films with thicknesses, $h_0$ = 3 nm, 5.5 nm and 11 nm, are described, which have been shown to lie in the deep, intermediate, and marginal spinodal region, respectively [4, 5].

In spinodal rupturing, the film roughens by spontaneous growth of the surface fluctuations, which can be regarded as the order parameter, and be decomposed into a sum of Fourier components according to mean-field theory [7]:

$$h(\vec{r},t) - h_0 = \sum_{\vec{q}} e^{\Gamma(\vec{q})t} A(\vec{q},0) \exp(i\vec{q}\cdot\vec{r}), \qquad (1)$$

where $i = \sqrt{-1}$, $\vec{r}$ is a 2D position vector on the film, $t$ is time, $A(\vec{q},0)$ is the amplitude of the mode with wave vector $\vec{q}$ at $t = 0$ and $\Gamma(\vec{q})$ is the growth rate. Implicit to eqn. (1) is the dynamic equation:

$$\frac{dA(\vec{q},t)}{dt} = \Gamma(\vec{q}) A(\vec{q},t). \qquad (2)$$

It can be shown that $\Gamma(\vec{q})$ is isotropic, and given by :

$$\Gamma(q) = \Gamma(q_m)[2(q/q_m)^2 - (q/q_m)^4], \qquad (3)$$

where $q \equiv |\vec{q}|$ and $q_m$ is the wave vector that maximizes $\Gamma(q)$, where

$$q_m = \left(\frac{G''(h_0)}{2\gamma}\right)^{1/2}. \qquad (4)$$

Here, $\gamma$ is the surface tension of PS (= 35 mJ/m$^2$) and $G(h_0)$ is the interfacial potential of the film that has initial uniform thickness equal $h_0$. For the present system, $G(h_0)$ usually assumes the van der Waals (vdW) potential [8-10] although alternative origins of $G(h_0)$ have been proposed [11, 12]. Since the time factor in eqn. (1) is exponential, the mode with $q = q_m$ quickly dominates, giving rise to a characteristic length = $2\pi/q_m$ in the rupturing morphology, and a ring with radius $q_m$ in the 2D Fourier spectrum. For $q$ greater than a cut-off of $\sqrt{2}q_m$, $\Gamma(\vec{q}) < 0$ with which the mode decays with time. The decay stems from the fact that the surface energy ($= \gamma |A(q)|^2 q^2/2$) of those modes with $q > \sqrt{2}q_m$ is so large that the total energy needed to create the mode becomes positive definite [7]. Equations (1)-(4) are the major equations governing the mean-field phase properties of a spinodally unstable film.

Polystyrene homopolymer with a molecular weight of 2.3 kg/mol (M$_w$/M$_n$ = 1.07) was purchased from Scientific Polymer Products. Solutions of this polymer in toluene with different concentrations were prepared and spun-cast at 4000 rpm to produce PS films with the desired thickness. The SiO$_x$ layer on the substrate was prepared by wet oxidation as detailed in Ref. [13]. The thickness of the PS films and the $SiO_x$ layer were measured by ellipsometry. Topographic images of the films as they ruptured were acquired by a Seiko Instruments SPN3800 atomic force microscope (AFM) operated in the non-contact mode. The annealing temperature of the sample was chosen such that the rupturing rate was convenient for the measurement. By analyzing the temporal development of the radial-averaged Fourier amplitude, $A(q,t)$ ($\equiv \langle A(\vec{q},t) \rangle_{|\vec{q}|=q}$), at each $q$, $\Gamma(q)$ can be determined.

Figure 1(a) shows three topographic images of a 3 nm film demonstrating the morphological development of the film in its early stage of rupture. In here and the rest of the topographic images shown below, the bright regions represent protrusions and the dark regions represent indentations. It is apparent from Fig. 1(a) that the growing surface fluctuations are correlated, as confirmed by the formation of a ring in the 2D Fourier spectrum of the images after ~3500 s (inset of Fig. 1(b)) and the development of a peak in the Fourier spectrum (main panel of Fig. 1(b)). As seen from Fig. 1(b), $A(q)$ develops a peak at $q_m/2\pi \sim 11\,\mu m^{-1}$ in the early stage, which shifts to smaller $q$'s at later times, in keeping with coarsening that takes place when the fluctuations have grown to make the higher-order terms significant [14]. Shown in Fig. 1(c) are displayed the time variations of the peak height of $A(q)$, $A_{max}$, and the depth, $d_h$, of five randomly selected holes as marked in Fig. 1(a). As seen, $A_{max}(t)$ increases exponentially with time initially, consistent with eqn. (1), but slows down notably after $t > 2.5 \times 10^4$ s. Interestingly, the evolution of $d_h$, which is essentially identical for all five holes, seems to follow that of $A_{max}$. The plateau in $d_h(t)$ commencing near $t = 2.5 \times 10^4$ s can be understood from $d_h$'s reaching $h_0$ whereupon its growth is hindered by the bottom substrate.  The exact correspondence between $d_h(t)$ and $A_{max}(t)$, and the simultaneity in the growth of all five randomly selected holes strongly suggests that the deepening of the holes had come from the amplification of the spinodal surface fluctuations.

Displayed in Fig. 2 are corresponding results obtained from a rupturing 5.5 nm film.  In contrast to the 3 nm film, the rupturing morphology of the 5.5 nm film is dominated by irregularly scattered holes that emerge at random times (Fig. 2(a)).  Nevertheless, the Fourier spectrum (Fig. 2(b)) still clearly shows a peak in about $6000\,s$.  The initial peak position, occurring at $q_m/2\pi \sim 2.5\,\mu m^{-1}$, does not correspond to any obvious length scales pertinent to the holes in the respective images.  Compared to the value of $q_m/2\pi \sim 11\,\mu m^{-1}$ found with the 3 nm film, the value obtained here is in reasonable agreement with the $q_m \sim h_0^{-2}$ scaling predicted by eqn. (4) assuming the prevalent form of $G(h_0) \sim h_0^{-2}$, which is appropriate for non-retarded vdW interactions [8-10] and confined phonon fluctuations [12].  In Fig. 2(c), $d_h$ of the five randomly selected holes shown in Fig. 2(a) demonstrates growth rates close to that of the spinodal fluctuations initially, but jump into full-thickness holes at sporadic times.  The random nature of the jump-in of the holes and a calculation [5] showing the opening of the holes could involve activation barriers comparable to $k_B T$ has led us to postulate that the holes are due to thermal nucleation [5].  Fig. 2(c) also shows that $A_{max}$ vs. $t$ displays an upturn near the time when the initial jump-in of holes takes place.  An upturn in $A_{max}$ vs. $t$ is inconsistent with the mean-field theory [7, 14].  As Fig. 2(a) shows and has always been found, $q_m$ decreases with $t$ at later times. Since the growth of surface waves with longer wavelengths necessarily involves the transport of more masses and hence must be slower, the growth of $A_{max}$ through spinodal dynamics must only slow down with $t$ as demonstrated by the 3 nm film.

In Figs. 3(a) and (b) are plotted the growth rate function, $\Gamma(q)$, of the 3 nm and 5.5 nm films in different time zones as defined in Figs. 1(c) and 2(c), respectively. The solid lines are fits to eqn. (3). For the 3 nm film, only data in time zone I display a good fit. Fits to the data in time zones II and III are marginal. Moreover, the data in time zones II and III display a short wavelength cut-off that is noticeably smaller than the mean-field prediction. Shifting of the short wavelength cut-off to a smaller value is commonplace and well documented [14, 15]. It arises from stochastic thermal agitations that help provide the energy required to excite the short-wavelength modes with $q > \sqrt{2}q_m$ [15]. For the $\Gamma(q)$ data of the 5.5 nm film, again only the data obtained in time zone I fit well to the model. The data obtained in time zone II, fitting marginally to eqn. (3), demonstrate growth rates that are way above those in time zone I, which is consistent with the upturn of $A_{max}(t)$ found in Fig. 2(c). However, the position of the peak is essentially unchanged. Independence of the growth rate on the position of $q_m$ is a strong indication that the growth dynamics in time zone II is not spinodal. The $\Gamma(q)$ data obtained in time zone III is essentially flat, showing no peak, nor a short wavelength cut-off, which are also unlike spinodal.

We examine a still thicker film with thickness equal 11 nm. Shown in Fig.

4(a) are three topographic images of the film representing its rupturing. As seen, there are growing fluctuations on the film surface from the beginning; at $t \sim 2700\,\text{s}$, the first hole appeared whereupon more holes emerged and increased in size with time. The data of $A_{\max}$ vs. $t$ (Fig. 4(b)) reveal additional details. The growth of the surface fluctuations actually slows down after $\sim 600\,\text{s}$ but shows an upturn later at $t \sim 2500\,\text{s}$, which approximately coincides with the time when the first hole appeared. To see if the upturn is connected with the emergence of the holes, we carefully select an area $(28 \times 28\,\mu\text{m}^2)$ inside the whole image $(48 \times 48\,\mu\text{m}^2)$ where no holes appear and reanalyzed $A_{\max}$ vs. $t$. The result is plotted as open circles in Fig. 4(b), where the upturn clearly goes away. Before the onset of the upturn, $\Gamma(q)$ (Fig. 5(a), time zones I and II) and $|A(q)|$ (Figs. 5(b) and (c)) of the whole image and the no hole area look alike. After the upturn in each of the time zones III and IV, $A_{\max}$ increases by 10 times and $q_m$ reduces by about 2 times in the whole image whereas $A_{\max}$ increases by only 2 times and $q_m$ reduces by only 20% - 30% in the no hole area. Immediately after the upturn in time zone III, $\Gamma(q)$ from the whole image greatly exceeds that in time zone II although the value of $q_m$ decreases by almost 3 times. Furthermore, the short-wavelength modes demonstrate substantial growth showing no sign of a cut-off. This is markedly different from the data from the no hole area where a cut-off in the growth rate is clearly apparent. The contrast with the presence of a cut-off is readily perceivable by comparing the Fourier spectra of the no hole area with those of the whole image (Figs. 5(a) and (b)). For the no hole area, little change in $|A(q)|$ takes place in the large $q$ region, but for the whole image, the growth of $|A(q)|$ is substantial over the entire $q$ range.

The influence of stochastic thermal noise on the dynamics of spinodal decomposition has been treated [14, 15]. By adding in a random force term of $q^2 \theta(\vec{q},t)$ (where $<\theta(\vec{q},\varsigma)\theta(\vec{q},\xi)> = |\theta(\vec{q})|^2 \delta(\varsigma-\xi)$) to the dynamic equation like eqn. (2), Cook showed that the solution for $|A(q,t)|^2$ is modified to [15, 16]:

$$|A(q,t)|^2 = \left\{|A(q,0)|^2 + \frac{q^4 |\theta(\vec{q})|^2}{\Gamma(q)}\right\} \exp[2\Gamma(q)t] - \frac{q^4 |\theta(\vec{q})|^2}{\Gamma(q)}. \qquad (5)$$

When $\Gamma(q)$ is negative, eqn. (5) describes how a system out of equilibrium is restored to the equilibrium spectral power density, $|A(q,\infty)|^2$, which can be identified with $-q^4 |\theta(\vec{q})|^2 / \Gamma(q)$ if $\Gamma(q) < 0$, by the thermal noise [15]. This force from the thermal noise causes the apparent growth rate, $\Gamma_{apparent}(q) \equiv (1/2t)\ln|A(q,t)/A(q,0)|^2$ to be bigger than the intrinsic growth rate, $\Gamma(q)$. Based on eqn. (5), $\Gamma(q)$ can be deduced from the measurement of $|A(q,t)|^2$ by

$$\Gamma(q) = \frac{1}{2t} \ln\left[\frac{|A(q,t)|^2 - I_{thermal}(q)}{|A(q,0)|^2 - I_{thermal}(q)}\right], \qquad (6)$$

where $I_{thermal}(q) \equiv -q^4 |\theta(\vec{q})|^2 / \Gamma(q)$. But when $\Gamma(q) > 0$ (as in spinodally unstable systems and $q < \sqrt{2}q_m$), $I_{thermal}(q)$ is negative and its physical significance of being $|A(q,\infty)|^2$ must cease. Nevertheless, the magnitude of $I_{thermal}(q)$ should still be determined by $|A(q,\infty)|^2$ since its origin with the random thermal noise, whose presence always facilitates the system's return to equilibrium, is unaltered. Using eqn. (6) and assuming that $|A(q,0)/A(q,\infty)|^2 \ll 1$, one finds:

$$\frac{\Gamma_{apparent}(q)}{\Gamma(q)} \cong \frac{1}{t'} \ln \left[ \mp \left| \frac{A(q,\infty)}{A(q,0)} \right|^2 (1 - \exp(\pm t')) \right], \tag{7}$$

where the upper (lower) sign is for $\Gamma(q), \Gamma_{apparent}(q) > 0$ ($< 0$) and $t' \equiv 2|\Gamma(q)|t$. By adopting the upper sign, $t' = 1$, and $|A(q_m,0)/A(q_m,\infty)|^2 \approx [h_{rms}(t_{upturn})/h_{rms}(\infty)]^2 \approx [h_{rms}(t_{upturn})/h_0]^2$ in eqn. (7), where $t_{upturn}$ is the time when the upturn in $A_{max}$ appears and $h_{rms}(t)$ is the rms roughness of the film at time $t$, we estimate that $\Gamma_{apparent}(q_m)/\Gamma(q_m) = 6.0$ and $8.9$ for the $h_0 = 5.5$ and $11$ nm films, respectively. The parameters involved in this calculation are shown in Table 1. This result can very well account for the noted factor of 3 rise in $\Gamma(q_m)$ after the upturn (Figs. 3(b) and 5(c)). We remark that the effect of thermal fluctuations on the growth of the spinodal unstable modes treated here is different from that by Cook [15, 16] who took $|A(q,\infty)|^2 = k_B T /(G''(h_0) + \gamma q^2)$ where the symbols, for the present problem, have the same meanings as in eqn. (4). The form of $|A(q,\infty)|^2$ used by Cook, deduced by the equi-partition law $|A(q,\infty)|^2 \sim k_B T / \chi(q)$ (where $\chi(q)$ is the $q$th Fourier component of the susceptibility of the system), is the *instantaneous* equilibrium spectral power density near $t = 0$ [17]. But as $|A(q,t)|^2$ grows large with time, non-linearity causes this instantaneous "$|A(q,\infty)|^2$" to change, inasmuch as it does in causing coarsening. Cook's result, consistent with the numerical result of Langer et al. [14], provides excellent account for the commonly found deviations between the measured $\Gamma(q)$ and the mean-field prediction, namely both the cut-off wavevector and the growth rate of the modes with $q$ far removed from $q_m$ are larger than expected, which are also evident in the 3 nm film (Fig. 3(a)). These deviations appear right from the beginning of the spinodal process. On the other hand, the deviations due to the upturn appear only after some time delay, suggesting the mechanism instigating the upturn necessitates the system to overcome some energy barriers. This characteristic distinctively matches with that of thermal nucleation.

In conclusion, we have observed unconventional deviations from the mean-field theory in the growth pattern of spinodal surface fluctuations of polymer films – notably the upturn in $A_{\max}(t)$, which we show arises from thermal nucleation. Since our discussions are general, the same irregularities should be present in any phase separation processes where the thermal noise induces large-amplitude order parameter fluctuations comparable in size to those at equilibrium. We recognize that this condition coincides with the Ginzburg criterion suggested by Binder [1] for the mean-field theory to be invalidated in the spinodal region. Result of this work shows that the large-amplitude fluctuations occurs by way of homogeneous nucleation.

This work is supported by the Research Grant Council of Hong Kong through the projects HKUST6070/02P and 603604.

* To whom correspondence should be addressed. E-mail: phtsui@ust.hk.

[1] K. Binder, *Phys. Rev. A* **29**, 341 (1984).

[2] W. Klein and C. Unger, *Phys. Rev. B* **28**, 445 (1983).

[3] Z. –G. Wang, *J. Chem. Phys.* **117**, 481 (2002).

[4] B. Du, F. Xie, Y. Wang, Z. Yang, and O.K.C. Tsui, *Langmuir* **18**, 8510 (2002).

[5] O. K. C. Tsui, Y. J. Wang, H. Zhao, and B. Du, *Eur. Phys. J. E* **12**, 417 (2003).

[7] A. Vrij and J. Th. G. Overbeek, *J. Am. Chem. Soc*. **90**, 3074 (1968).

[8] G. Reiter, *Phys. Rev. Lett*. **68**, 75 (1992).

[9] R. Xie, A. Karim, J. F. Douglas, C. C. Han, and R. A. Weiss, *Phys. Rev. Lett*. **81**, 1251 (1998).

[10] R. Seemann, S. Herminghaus, K. Jacobs, *Phys. Rev. Lett*. **86**, 5534 (2001).

[11] H. Zhao, Y. J. Wang, and O. K. C. Tsui, *Langmuir* **21**, 5817 (2005), introduction.

[12] M. D. Morariu, E. Schäffer, and U. Steiner, *Phys. Rev. Lett*. **92**, 156102 (2004).

[13] X. P. Wang, X. Xiao, and O. K. C. Tsui, *Macromolecules* **34**, 4180 (2001).

[14] J. S. Langer, M. Bar-on and H. D. Miller, *Phys. Rev. A* **11**, 1417 (1975).

[15] H. E. Cook, *Acta Metallurgica* **18**, 297 (1970).

[16] We found in Ref. [15] that a negative sign was missing from the exponential in the equation after eqn. (17) and all the terms involving $Q(\vec{k})$ between eqn. (18) and (21).

[17] We remark that for $q < \sqrt{2} q_m$, the quantity used by Cook in Ref. [15] for $|A(q,\infty)|^2$ is negative and hence should actually not be taken as the equilibrium spectral power density (see the discussion given in the text for the equivalent quantity $I_{\text{thermal}}(q)$ used in this paper).

| $h_0$ (nm) | $h_{\text{rms}}(t_{\text{upturn}})/h_0$ | $\Gamma_{\text{apparent}}(q_m)/\Gamma(q_m)$ (Theory) | $\Gamma(q_m)_{\text{after upturn}}/\Gamma(q_m)_{\text{before upturn}}$ (Experiment) |
|---|---|---|---|
| 3 | 0.067 (at $t=0$) | 5.9 | No upturn |
| 5.5 | 0.064 | 6.0 | 3 |
| 11 | 0.022 | 8.2 | 3 |

**Table 1** Parameters involved in the calculation of $\Gamma_{\text{apparent}}(q_m)/\Gamma(q_m)$, which is compared to the ratio of the observed $\Gamma(q_m)$ before and after the upturn (last column).

FIGURE CAPTIONS

**Fig 1. (a)** AFM topographic images from a rupturing 3 nm film at different $t$ indicated. The scale bar on the first image applies to all images. **(b)** $|A(q)|$ of the same film plotted versus $q/2\pi$ at various $t$ (indicated for the spectra drawn in solid lines). (Inset) A typical 2D Fourier spectrum of the film after a peak appears in $|A(q)|$. Darker colors represent larger signals. **(c)** Simultaneous plots of $A_{max}$ and holes depth versus time. The analyzed holes are those labeled #1 to #5 in (a).

**Fig 2. (a)** AFM topographic images from a rupturing 5.5 nm film at different $t$ indicated. The scale bar on the first image applies to all images. **(b)** $A(q)$ of the same film plotted versus $q/2\pi$ at various $t$ (indicated for the spectra drawn in solid lines). **(c)** Simultaneous plots of $A_{max}$ and holes depth versus $t$.

**Fig. 3.** The $\Gamma(q)$ function of a 3 nm **(a)** and a 5.5 nm **(b)** film in different time zones defined in Fig. 1(c) and 2(c), respectively. Solid lines are fits to eqn. (3).

**Fig. 4. (a)** AFM topographic images obtained from a rupturing 11 nm film at different times as indicated. The area framed by dashed lines are selected to contain no hole and are termed "no hole area" in the text. **(b)** Plots of $A_{max}$ versus $t$ from data of the whole image (solid squares) and the no hole area (open circles) shown in (a).

**Fig 5. (a)** The $\Gamma(q)$ function deduced from the whole image and the no hole area of the same 11 nm film at different time zones in Fig. 4(b). Solid lines are fits to eqn.

(3). **(b)** $|A(q)|$ versus $q/2\pi$ of this film from the whole image and **(c)** the no hole area at various $t$ (indicated for the spectra drawn in solid lines)

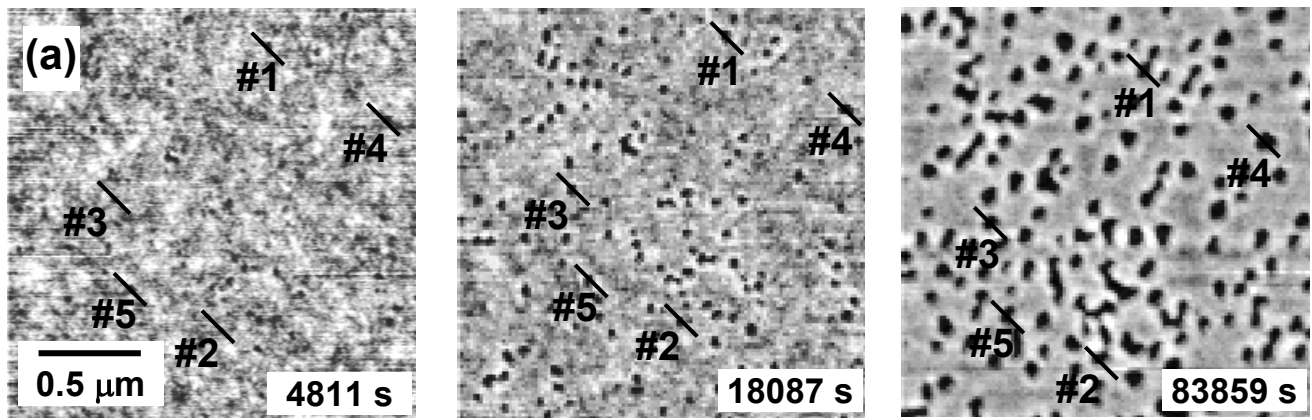
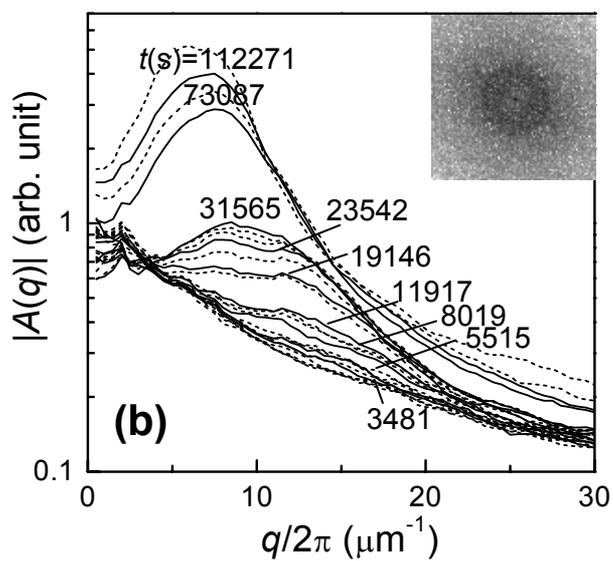
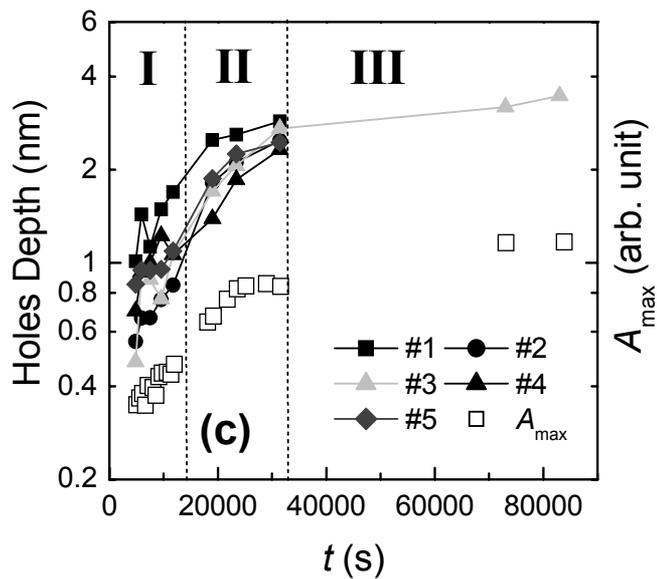

Fig. 1

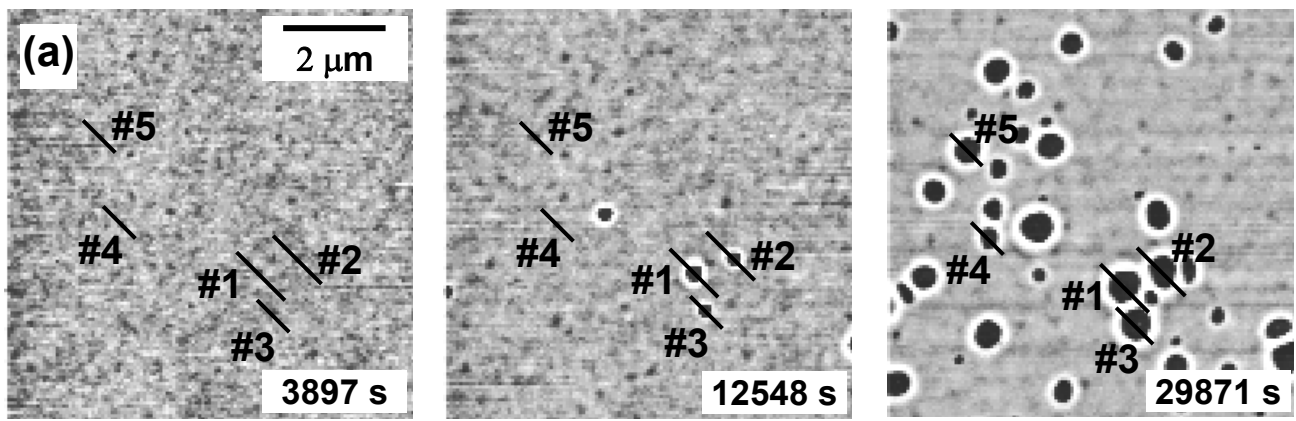
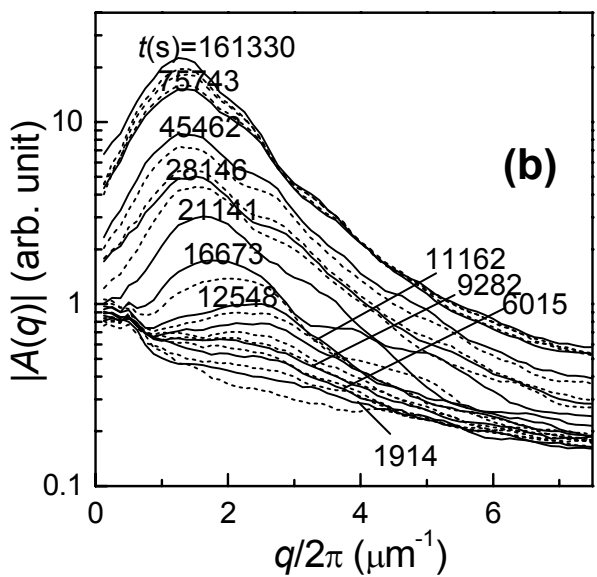
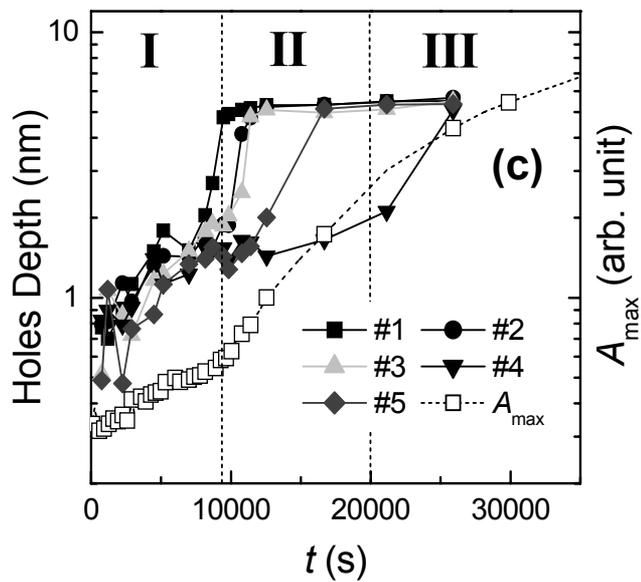

Fig. 2

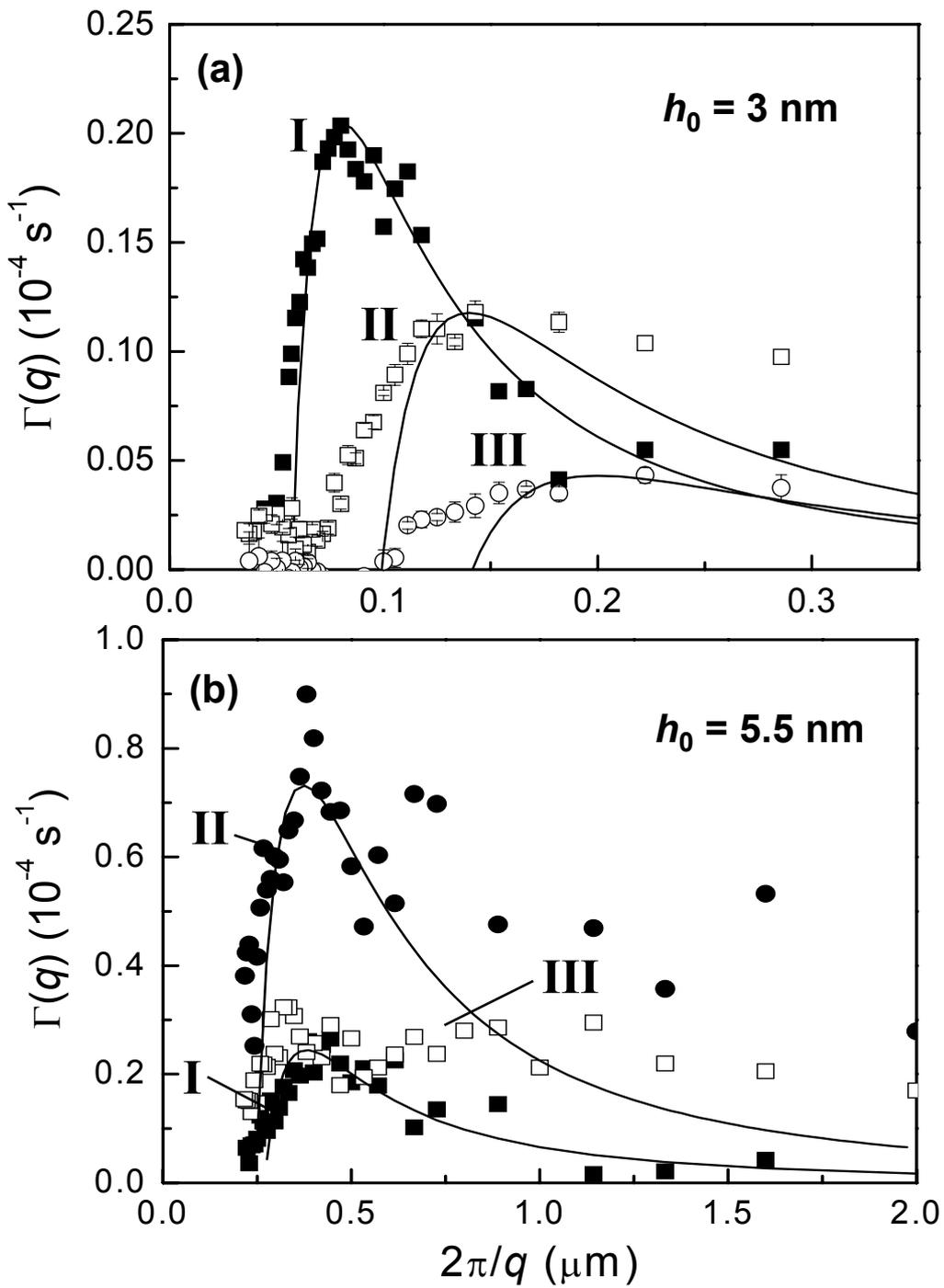

Fig. 3

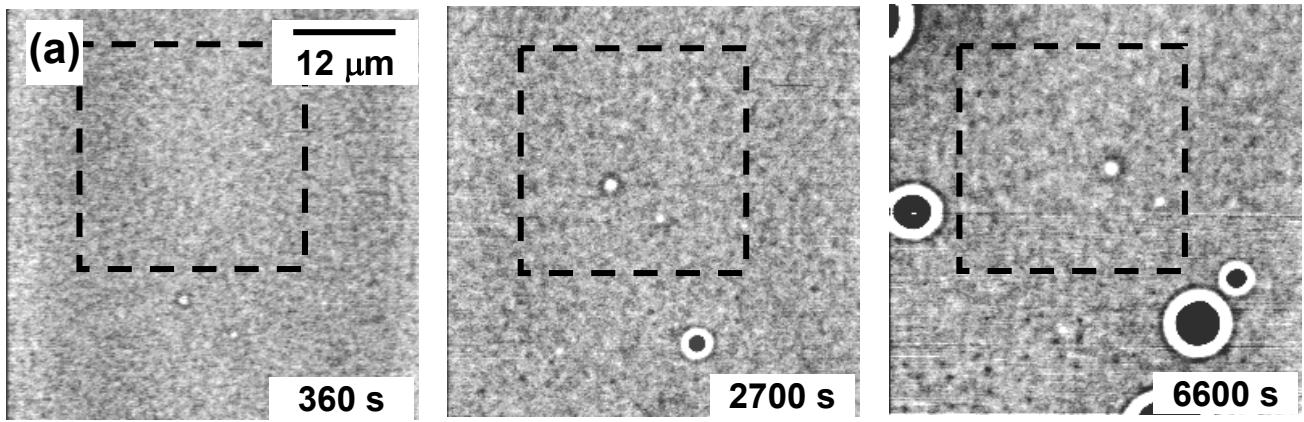
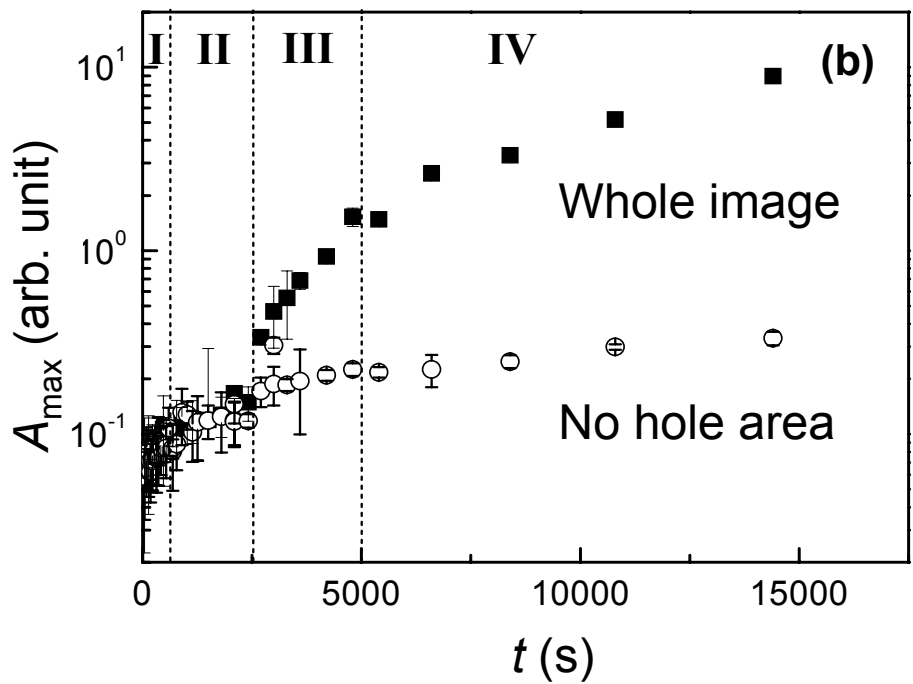

Fig. 4

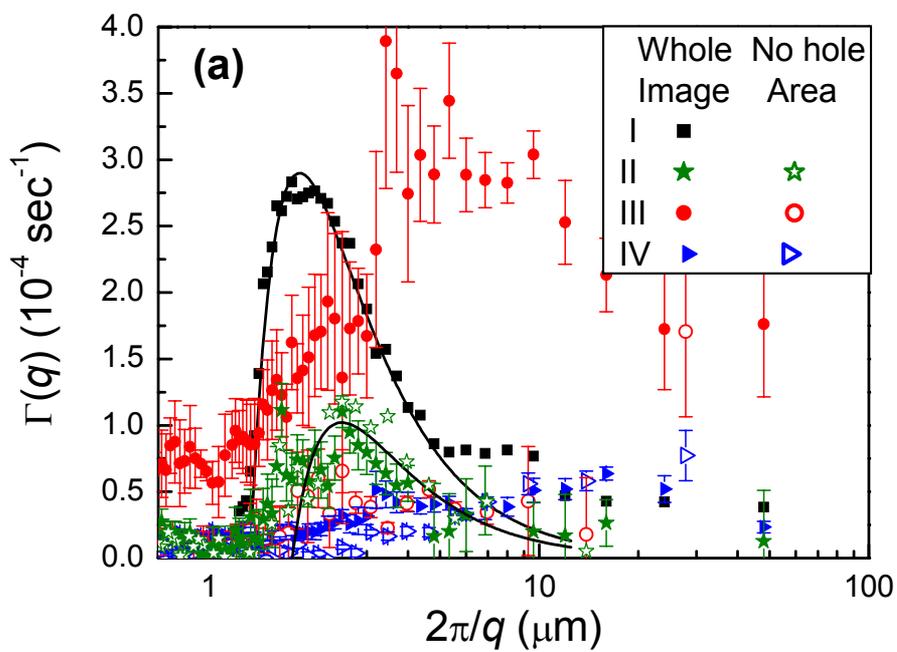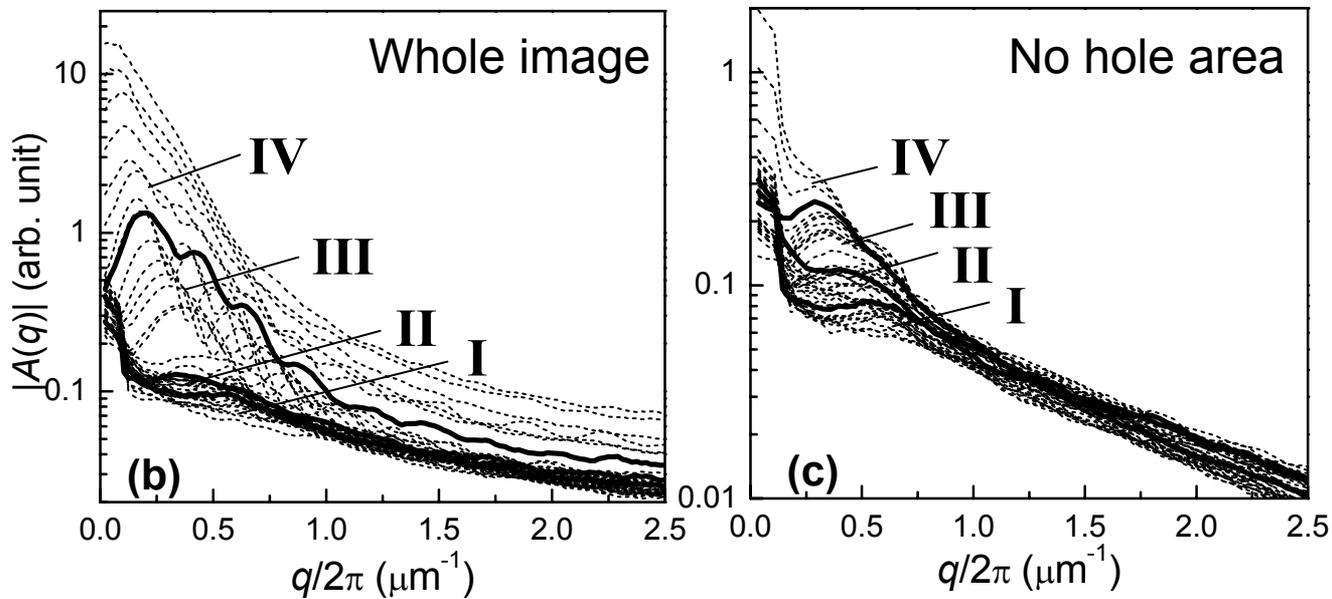

Fig. 5